# Film structure of epitaxial graphene oxide on SiC: Insight on the relationship between interlayer spacing, water content, and intralayer structure

*S. Zhou, S. Kim, E. Di Gennaro, Y. Hu, C. Gong, X. Lu, C. Berger, W. de Heer, E. Riedo, Y. J. Chabal, C. Aruta, and A. Bongiorno[*]*

S. Zhou, Prof. Angelo Bongiorno
  School of Physics, Georgia Institute of Technology,
  Atlanta, Georgia 30332-0430, USA
  and
  Chemistry and Biochemistry, Georgia Institute of Technology,
  Atlanta, Georgia 30332-0400, USA
  E-mail: angelo.bongiorno@chemistry.gatech.edu

Prof. S. Kim
  Department of Applied Physics, Hanyang University, Ansan 426-791, South Korea

Dr. E. Di Gennaro, Dr. C. Aruta
  CNR-SPIN, Dipartimento di Ingegneria Civile e Ingegneria Informatica, University
  of Roma Tor Vergata, 00133 Rome (Italy) and Dipartimento di Scienze Fisiche,
  Monte S. Angelo, 80126 Naples (Italy)

Y. Hu, X. Lu, Dr. C. Berger, Prof. W. de Heer, Prof. E. Riedo
  School of Physics, Georgia Institute of Technology,
  Atlanta, Georgia 30332-0430, USA

C. Gong, Prof. Y. J. Chabal
  Department of Materials Science and Engineering, The University of Texas at
  Dallas, Richardson 75080, Texas, USA

Dr. C. Berger
  School of Physics, Georgia Institute of Technology,
  Atlanta, Georgia 30332-0430, USA;
  Centre National de la Recherche Scientifique - Institut Néel, Grenoble, B.P. 166,
  38042, France

Keywords: Graphene Oxide, Multilayer Epitaxial Graphene, XRD, ATR-IR, DFT calculations

# Abstract

Chemical oxidation of multilayer graphene grown on silicon carbide yields films exhibiting reproducible characteristics, lateral uniformity, smoothness over large areas, and manageable chemical complexity, thereby opening opportunities to accelerate both fundamental understanding and technological applications of this form of graphene oxide films. Here, we investigate the vertical inter-layer structure of these ultra-thin oxide films. X-ray diffraction, atomic force microscopy, and IR experiments show that the multilayer films exhibit excellent inter-layer registry, little amount (<10%) of intercalated water, and unexpectedly large interlayer separations of about 9.35 Å. Density functional theory calculations show that the apparent contradiction of "little water but large interlayer spacing in the graphene oxide films" can be explained by considering a multilayer film formed by carbon layers presenting, at the nanoscale, a non-homogenous oxidation, where non-oxidized and highly oxidized nano-domains coexist and where a few water molecules trapped between oxidized regions of the stacked layers are sufficient to account for the observed large inter-layer separations. This work sheds light on both the vertical and intra-layer structure of graphene oxide films grown on silicon carbide, and more in general, it provides novel insight on the relationship between inter-layer spacing, water content, and structure of graphene/graphite oxide materials.

# 1. Introduction

Graphene oxide (GO)[1-3] holds great promise for a variety of applications, including supercapacitors,[4,5] optical devices,[6] and mechanical actuators.[7-9] GO is a complex nonstoichiometric and hygroscopic material, and understanding and controlling its structural and physical chemical properties are matter of intense research.[2,7,9-17] Synthesized for the first time in 1859,[18] today GO is generally obtained via the so-called modified Hummers' methods.[3,13,19] These methods involve chemical oxidation of graphite and several solution processing steps in order to exfoliate, filtrate, and deposit graphene oxide flakes on a substrate. The resulting paper-like material consists of a disorganized stacking of oxidized carbon platelets[3,13,19] which, due to their hydrophilic nature, are prone to adsorb and incorporate water molecules into the galleries of the film.[20] X-ray diffraction measurements of conventional GO in the form of powder or thick films show that the spacing between the oxidized carbon layers increases from 5-6 Å to about 8-12 Å by increasing the water content in the film up to about 25 wt%,[11,12,14] a concentration reachable by hydrating GO for days at a relative humidity of 100%.[11,12,14] These experimental observations are explained in terms of the well-accepted picture that water molecules intercalate the lamellar film pushing the GO layers apart, and that this effect increases for increasing the water content in the film. While hydration behavior of GO and dynamics of intercalated water are well-documented[11,12,14], the relationship between interlayer spacing, water content, and GO film structure remains elusive and only intuitively understood.

In this work, we investigate the inter-layer structure of graphene oxide films obtained by chemical oxidation of multilayer epitaxial graphene (EG) grown on silicon carbide.[17] These oxide films – henceforth referred to, for convenience, as EGO – are synthesized directly on EG/SiC chips through the use of a "mild" Hummers process without requiring any exfoliation, filtration or transfer processes (Fig. 1(a)). Recent studies have shown that this synthesis strategy and the resulting EGO films present several advantages over the conventional synthesis method and GO, such as the rapidity of the synthesis process – between 1 and 2 hours (see Methods section), improved reproducibility of the film properties, the extended uniformity and smoothness of the films,[17,20] the absence of edges and holes, and the fact that the films are supported by SiC chips which may facilitate the fabrication of high-quality electronic devices,[21,22] sensors,[23] and supercapacitors.[4,5] While in recent studies, experiment[17] and computations[24,25] were used to study the intra-layer chemical structure of EGO, here we combine experiments and density functional theory (DFT) calculations to address the vertical structure of the EGO multilayer films and elucidate the relationship between inter-layer separation, intra-layer spatial distribution of oxygen functionalities, and water content. In particular, our X-ray diffraction (XRD) and atomic force microscopy (AFM) experiments show that multilayer EGO retains the order of the carbon lattice, number of layers, and spatial coherence lengths of the pristine EG films, and that in ambient conditions, the inter-layer separation of the well-registered graphene oxide layers is 9-10 Å. X-ray photoelectron spectroscopy (XPS) shows that EGO films present O/C ratios of about 0.38. Furthermore, XRD, XPS, and IR measurements concur in showing that the EGO films contain minimal concentrations of water, estimated to be no larger than 10 wt%, thereby demonstrating that EGO combines large inter-later separations and little amounts of water intercalated in the film, a behavior contrasting that one exhibited by conventional GO. Our experimental results are explained on the basis of models derived from DFT computations. In particular, our computations show that the experimental results are consistent with a multilayer film formed by carbon layers presenting – at the nanoscale – a non-homogenous oxidation, where non-oxidized and highly oxidized nano-domains coexist, and where only a few water molecules trapped between the oxidized regions of the stacked layers are sufficient to account for the measured large inter-layer distances of 9-10 Å. The proposed film structure for EGO and the occurrence of water molecules bridging and pinching together the oxidized layers of EGO are consistent with the fact that our EGO films exhibit high resistance towards exfoliation and thus that they retain the number of carbon layers of the pristine EG films in spite of the aggressive chemical oxidation method and rinsing steps used during their synthesis.

## 2. Results and Discussion

### 2.1 Experiments

The EGO films were obtained by a "mild" Hummers oxidation of EG films (Figure 1(a)) consisting of 9 to 12 graphene layers grown epitaxially on the C-face of SiC (000-1) substrates.[26] At variance with the conventional process used to oxidize graphite and produce GO, here the duration of the whole synthesis process is only 2 hours, more details are reported in the Methods part. Here, we study EGO films aged for more than 1 month in ambient conditions at a relative humidity (RH) of about 50%. The area of the films was either 6x4 mm$^2$ or 10x10 mm$^2$. Chemical compositions derived from our XPS spectra of three EGO samples are reported in Table 1; details of this analysis are reported in the Supplementary Information. A full discussion and details about EG films and the oxidation process, experiments, and DFT calculations are reported in the Methods section and Supplementary Information (SI).

To measure interlayer spacing and assess film quality, we first used XRD. Figure 1(b) shows the ω-2θ XRD spectrum of a 11±1-layer EGO film, along with the spectra obtained from a pristine 11±1-layer EG film and the bare SiC substrate. Our XRD measurements show that the main peak of the 4H-SiC substrate occurs at 2θ=35.65 deg, while the sharp peaks at 2θ=8.78, 17.61 and 26.54 deg correspond to quasi-forbidden reflections. The XRD peaks arising from the 11-layer EGO and EG films are observed at 2θ=9.46 and 2θ=26.42 deg, corresponding to interlayer distances of 9.35 and 3.38 Å, respectively. These two peaks are normalized to the maximum peak intensity and confronted in Fig. 1(c). This comparison shows that the EG and EGO films composed of a similar number of layers yield diffraction peaks with similar values of the full-width-at-half-maximum (FWHM), indicating that the vertical coherence of the layered structure of EGO is comparable to that one of EG. The rocking curves (ω-scan) in Fig. 1(d) corroborate this result. These measurements, sensitive to interlayer orientation, show in fact that the peaks associated to EGO, EG, and SiC exhibit similar values of the FWHM, arising in all cases mostly from instrumental broadening. Overall, our XRD measurements show that the EGO and EG films have comparable vertical interlayer registry, thereby indicating that the mild Hummers oxidation step employed in this work to produce EGO allows to transfer the excellent film quality of EG to EGO. Furthermore, the XRD measurements show that the distance between the graphene oxide layers in EGO is as large as 9.35 Å. These results are reproducible and little dependent on the thickness of the multilayer EGO films. XRD measurements of four different EGO films led to interlayer distances between 9.35 and 10.03 Å (SI).

Experimental[7, 11, 12, 14] and computational[7, 9] studies of conventional GO obtained from Hummers oxidation of graphite – followed by exfoliation, filtration, transfer, and deposition steps – concur in showing that the interlayer separation in GO relates to the amount of water intercalated in-between the GO platelets. In particular, both experimental[7, 14] and computational[7, 9] studies show that to achieve inter-layer distances larger than 9 Å, GO ought to incorporate large and detectable quantities of water, amounting to more than 20 wt%. To probe the presence of intercalated water in our EGO films, we used different strategies. First, we annealed EGO for 1 hour at, sequentially, 100, 120, 140, and 160°C. After each annealing step, we cooled down the EGO films to room temperature and performed XRD measurements (see Figure S3 in SI). These thermal treatments and experiments showed that EGO is thermally stable up to 140°C, and that after annealing EGO at this temperature, the interlayer spacing dropped to only 8.46 Å. It is to be noted that similar temperature-resolved XPS measurements show a drop of number of C-O bonds in EGO films annealed at around 120-140°C, with a consequent drop of the C:O ratio to about 0.3. These observations indicate that at temperatures larger than 120°C the EGO films start to reduce and lose oxygen, and that in spite of this the interlayer spacing remains larger than 8 Å. Second, we carried out XRD measurements of an EGO sample before and after drying it in a $N_2$ atmosphere over one night. This drying treatment led to no significant changes in the XRD spectra (Fig. 1(e)), indicating that the large inter-layer distance of 9.35 Å in EGO does not arise from the presence of large amount of water intercalated in the film. Third, we carried out attenuated total reflectance (ATR) IR measurements (Figure 2(a)). In these measurements, detection of water is usually best accomplished in the region of the $H_2O$ scissor mode that is not subject to strong H-bonding shift. The OH stretch region is more difficult to analyze because of water fluctuation in the spectrometer and on the liquid $N_2$-cooled IR detector. In this case, however, the SiC itself has some weak phonon overtone absorption in the 1500-1700 cm$^{-1}$ region, requiring some care in analyzing the IR data in the region of the $H_2O$ scissor mode (SI). Consequently, examination of the IR spectra in the 1500-1700 cm$^{-1}$ region confirms the attenuation of the SiC phonon modes due to thickness changes but cannot precisely quantify the water amount in the EGO films. Nevertheless, since the detection limit of water is approximately 1 monolayer (*i.e.*, ~$6\times10^{14}$ $H_2O$ molecules/$cm^2$) and there are ~10 layers in EGO (~$2.5\times10^{15}$ C atoms/$cm^2$/layer), our ATR-IR measurements set an upper limit for the amount of water in EGO of 10% (relative to the total amount of C). This estimate is in agreement with the that one extracted by analyzing the XPS spectra of EGO (see Table 1 and SI). Overall, these experiments indicate that the large inter-layer separation of 9.35 Å of EGO is not attributed to the presence of large quantities of water inside the films, but it is an intrinsic structural property of these ultra-thin oxide films.

To corroborate our XRD results, we used AFM to investigate both the lateral and vertical structure of the EGO films (Figures 2(b)-2(d)). In particular, our AFM topographic images show that the EGO films are homogeneous and exhibit lateral uniformity with typical height variations of 2.5 Å over the micrometer scale, and an average surface roughness at the nanometer scale of only 0.6 Å (Fig. 2(c)). The bright lines – corresponding to elevated heights – are visible in the AFM topographies of both the EGO and EG films (SI). These lines show the occurrence of pleats of the graphene layers, arising due to the different thermal expansion coefficients of graphene and SiC during the cooling of EG after its epitaxial growth at high temperature. More interestingly, AFM height profiles (Figure 2(c)) of an 11-layer EG film before and after oxidation, across a trench produced in the same EG film before oxidation, reveal that upon oxidation the films expand in the vertical direction by a factor between 2.5 and 3, a value in close agreement with the expansion of the interlayer distance from 3.38 to 9.35 Å detected by XRD. Thus, in agreement with our XRD measurements, these results show that mild Hummers oxidation of EG leads to EGO films preserving number of layers and the lateral and vertical uniformity of the un-oxidized EG films.

**2.2 DFT calculations**

To elucidate film structure and the origin of the large inter-layer distance in EGO, we used a DFT scheme complemented with semi-empirical corrections to account for London dispersion forces, the so-called DFT-D2 scheme.[27] This modeling scheme was recently used to interpret experimental XPS spectra of EGO and elucidate the intra-layer chemical structure of both as-synthesized EGO films and EGO films aged at room temperatures in ambient conditions.[17, 24, 25] In these previous computational studies, EGO was modeled as a stack of graphene layers hosting a homogeneous distribution of epoxide and hydroxyl species, and incorporating a fraction of $H_2O$ molecules (relative to the total amount of C) up to 10%.[17, 24] These types of model structures of EGO (see Figures 3 and 4(a)) exhibit interlayer distances ranging between 4.5 Å and 7.5 Å, depending on the oxidation level, distribution of oxygen functional groups, and water content.[17, 24] These results, as well as recent molecular dynamics studies of GO,[7, 9] show that graphene oxide layers presenting O:C ratios as high as 0.5 and exhibiting homogeneous distributions of oxygen functionalities on the carbon basal planes can account for the chemical features of GO observed by XPS, but they cannot explain the simultaneous occurrence in EGO of interlayer distances of 9.35 Å and water contents not exceeding 10%, as indicated by our XRD, IR, and AFM experiments (Figures 1 and 2, and SI).

To elucidate our experimental observations, we follow up the results of a recent computational study of GO,[25] showing that epoxide and hydroxyl species are prone to agglomeration and that ageing or annealing at moderate temperatures favor segregation phenomena and the formation of graphene oxide structures consisting of an interpenetrating network of highly oxidized and non-oxidized domains having nanometer dimensions.[25] A graphene oxide layer presenting a realistic composition – i.e. O:C=0.38 and fractions of epoxide and hydroxyl species equal to 0.11 and 0.27, respectively – and exhibiting such an inhomogeneous oxidation is schematized in Figure 4(d); this model structure was generated by simulating ageing phenomena in a homogeneously oxidized graphene oxide layer, as described in Ref.[25]. Here, we address the effect of an inhomogeneous structure of the graphene oxide layers on the interlayer separation of a multilayer EGO film. To this end, we considered model structures of EGO consisting of periodic stacks of graphene layers fully oxidized by either hydroxyl or epoxide groups, including increasing concentrations of water molecules, and presenting different stacking structures (Figs. 3(a) and 3(b) and SI). For each model structure, we used DFT-D2 to perform a full structural optimization and determine the zero-temperature interlayer spacing. The full set of calculations and results are reported in the SI, while the important insights derived from our computations are summarized in Figure 3.

Our DFT-D2 calculations show that graphene layers fully oxidized with epoxide species exhibit a smaller interlayer spacing than layers fully oxidized with hydroxyl groups, regardless the amount of $H_2O$ molecules intercalated between the functionalized layers (Figures 4(b) and 4(c)). In the former case, the interlayer spacing reaches a value of "only" 7.5 Å when the water content is 25%, while carbon sheets oxidized with hydroxyl groups attain separations of 8.6 Å when only 6.25% $H_2O$ is intercalated between the layers. This latter value is close to the one observed experimentally. In both cases, the interlayer spacing in EGO models containing 6.25% $H_2O$ differ significantly from that one assumed in dry conditions, increasing very little for increasing water concentrations. Trapping water molecules at oxidized regions of the multilayer graphene film is favored energetically. Figures 4(b) and 4(c) report the enthalpy difference per water molecule between the dry and hydrated EGO model structures of epoxide- and hydroxyl-only functionalizations of multilayer graphene, respectively. These relative enthalpy values derived from our DFT-D2 calculations show that there is an enthalpic driving force leading water molecules in EGO to more away from non-oxidized regions of a graphene layer – where they physisorb with enthalpic gains of about 0.03 eV – to accumulate at oxidized regions of the lamellar film. Interestingly, the enthalpic driving force to expand interlayer separation and trap water molecules is larger at oxidized domains of EGO rich in hydroxyl rather than epoxide groups; in both cases the enthalpy gain per water molecule increases for increasing the concentration of water molecules (Figure 4).

Overall, our DFT-D2 calculations support the idea of EGO films presenting the structure sketched in Figure 4(e). The graphene oxide layers present – at the nanoscale – a non-homogeneous oxidation consisting of highly oxidized areas – rich in hydroxyl groups – surrounded by nano-domains of non-oxidized graphene. According to this model, a few (<10%) water molecules forming strong hydrogen bonds with hydroxyl groups of nearest neighbor layers would thus be sufficient to have interlayer spacing up to 9 Å, thereby explaining our XRD, AFM, and IR experimental results.

## 3. Conclusions

We used XRD, AFM, and IR measurements and DFT-D2 calculations to investigate the film structure of EGO thin films synthesized by mild Hummers oxidation of multilayer epitaxial graphene on SiC. Our experiments show that the EGO films on SiC are uniform and homogeneous over the micrometer scale across the whole film area, that they retain the lateral order of the Carbon lattice, the number of layers, and film registry of pristine EG, and that they exhibit an interlayer spacing larger than 9.35 Å, in spite of the minimal concentration of water molecules intercalated into the multilayer structure. Our DFT calculations show that film structure of EGO consists – most likely – of a stack of layers exhibiting, at the nanoscale, phase separation in highly oxidized and non-oxidized graphene domains, as well as an inhomogeneous distribution – at the nanoscale – of the water molecules trapped in-between the oxidized regions of nearest neighbor layers. The highly oxidized regions are rich in hydroxyl species, they are prone to trapping water molecules, and thanks to the film planarity and its mesoscopic homogeneity, only a small amount (<10%) of water is required to achieve interlayer separations of 9.35 Å. Overall, this study provides new insight on the relationship between interlayer spacing, water content, and intra-layer structure of graphite/graphene oxide materials. Furthermore, this investigation shows that EGO presents unique film characteristics, including rapid and reproducible synthesis, extended uniformity, direct synthesis on SiC chips, and interlayer registry, that render these films suitable for high-quality electronic devices,[21, 22] sensors,[23] and supercapacitors,[4, 5, 28] as well as fundamental studies about the chemistry and structure of functionalized graphene.

# 4. Methods

## 4.1 Chemical (Hummers) oxidation of epitaxial graphene

To oxidize EG, we used the following method. Sodium nitride (250 mg) and sulfuric acid (98% concentrated 11.5 ml) were mixed in a beaker at room temperature. EG films (8-11 layers) 4x6 mm$^2$ were carefully placed into the mixed solution. The solution was stirred and cooled down to 4°C placing the beaker in an ice bath. Next, potassium permanganate (1.5 g) was added slowly into the beaker while shaking it. After adding all oxidation agents, the beaker was shaken until the solution was evenly mixed. The beaker then was moved to a 35°C water bath. The beaker kept shaken for a minute every 10 minutes to make sure the solution dissolved completely. After 30 minutes, the beaker was taken out of the water bath. Slowly, DI water (23 ml) was added into the solution. For the first 10 ml, DI water was added in drop by drop and after that 3~5 drops were added each time. The beaker was shaken during the entire adding procedure. The beaker was then left for 15 minutes with the top covered. Finally, 75 ml DI water and 1.5 ml hydrogen peroxide (3%) were added to the solution. The solution was stirred till it turned out to be transparent. EGO films were picked up from the solution and rinsed with DI water for 1 minute. The EGO samples were finally blow-dried by nitrogen gas. AFM topography images of a pristine EG films and the EGO film resulting from the aforementioned oxidation method are shown in Figure S1. The EGO films used in this study had a number of layers between 9 and 12. We have not observed significant changes in properties in this range of thicknesses. The experiments have been performed on several EGO samples, however when possible we report the data on specific EGO films for which we analyzed the corresponding EG film before oxidation.

**4.2 XRD and XPS measurements**

To obtain information about the vertical lattice spacing and the crystallographic quality of EGO films, we have performed X-ray diffraction (XRD) measurements at Cu $K\alpha$ wavelength in specular geometry. The spectra have been measured after the alignment of the SiC(000-1) substrate crystallographic planes. The geometry of these measurements is reported in Figure S2. The $\omega$-$2\theta$ XRD spectra of an 11-layer EGO film, a 12-layer EG film, and the bare SiC substrate are shown in Figure S2.

A commercial monochromatic thermo K-Alpha X-ray photoelectron spectrometer system (Thermo Scientific) is used for XPS characterization. We employ the software "XPSPEAK41" for the spectral data analysis. After a Shirley background subtraction, and correction for the different X-ray cross-sections using Scofield sensitivity factors, the C $1s$ spectra are fitted with three Gaussian-Lorentzian peaks with the constrain of equal full-widths at half-maxima and equal Gaussian-Lorentzian pre-factors. The three peaks are named in the text as $P_G$, $P_{GO}$, and $P_{carbonyl}$. A Gauss-Lorentzian function is chosen for the peak analysis since the material investigated here, multilayer graphene oxide, is nonmetallic. More details are found in Ref. 17.

**4.3 AFM and other measurements**

AFM topography images are obtained by a Veeco Nanoscope IV Multimode in contact mode. To measure the thickness of an EGO film after oxidation of EG, a razor blade is used to remove EG in some areas and to create a step between the EG film and the SiC substrate. Afterwards, a clean area of 1 µm$^2$ that includes a step feature was selected for imaging before and after oxidation (Figure 2 and S3) to examine the layer thickness variation due to the Hummers oxidation process. Histograms, obtained through the topography analysis using the WSxM software from the 1 µm$^2$ area, are also shown in Figure S3.

We measured the contact angle of EGO and graphene to be 48 ± 6 ° and 85 ± 4 °, respectively. AFM Kelvin probe microscopy shows that the surface contact potential difference between graphene and EGO is equal to 252 mV.

**4.4 IR measurements**

EGO is annealed in air at a series of temperatures up to 200°C in a Linkam FTIR 600 cooling/heating stage. Each annealing is done by increasing the temperature at the rate of 5°C/minute, holding at the target temperature for 1 hour, and cooling by shutting off the heater (estimated initial cooling rate of ~20oC/min). The sample is then measured by attenuated total reflectance infrared (ATR-IR) spectroscopy by pressing it against a germanium crystal with a weight of 56 oz., ensuring reproducibility and quantitative comparison between runs.

**4.5 DFT calculations**

Density functional theory (DFT) calculations were carried out by using the *PWscf* code contained in the QUANTUM-Espresso package.[27] We used an plane-wave energy cutoff of 120 Ry to represent the Kohn-Sham wavefunctions, norm-conserving pseudopotentials for all atomic species,[29] and the generalized-gradient-approximation exchange-correlation functional of Perdew, Burke, and Ernzerhof.[30] We used 3x3x3 and 3x3x1 Monkhorst-Pack meshes centered in the Γ-point to sample the Brillouin zone of supercells including one and two oxidized graphene layers, respectively. To describe the London dispersion forces between the graphene oxide layers, we used the semi-empirical DFT-D2 scheme proposed by Grimme.[31]

# Supporting Information

Supporting information is available from the Wiley Online Library or from the corresponding author.

# Acknowledgements

S.Z., S.K., C.-Y.C., X.L., H.-C.C., E.R., and A.B. acknowledge the support of the National Science Foundation (NSF) grants CMMI-1100290 and DMR-0820382. Y.H., C.B. and W.d.H. acknowledge the support of NSF grant DMR-0820382. A.B. acknowledges the support of the Samsung Advanced Institute of Technology (SAIT) and NSF grant CHE-0946869. E.R. acknowledges the support of the NSF grant DMR-0706031 and the Office of Basic Energy Sciences of the US Department of Energy (DE-FG02-06ER46293). Y.J.C. and C.G. acknowledge the support of the Office of Basic Sciences of the US Department of Energy (DE-SC001951).

| Fractions | EGO$_1$ (70 days) | EGO$_2$ (75 days) | EGO$_3$ (>365 days) |
|---|---|---|---|
| P$_{oxygen}$ (O/C) | 0.38 | 0.41 | 0.37 |
| P$_{GO}$ (C-O/C) | 0.45 | 0.45 | 0.37 |
| P$_{epoxide}$ (C-O-C/C) | 0.10 | 0.06 | 0.06 |
| P$_{hydroxyl}$ (C-OH/C) | 0.24 | 0.32 | 0.24 |
| P$_{carbonyl}$ (C=O/C) | 0.04 | 0.03 | 0.06 |
| P$_{Water}$(H$_2$O/C) | <0.14 | <0.17 | <0.13 |

**Table 1.** Fractions relative to total amount of C of O, C-O bonds, epoxide groups, hydroxyl species, carbonyl groups, and water molecules three EGO films aged in air for 70, 75, and more than 365 days. These fractions are derived by analyzing our XPS spectra[17, 24]. The fractions of water molecules in the films correspond to upper limits.

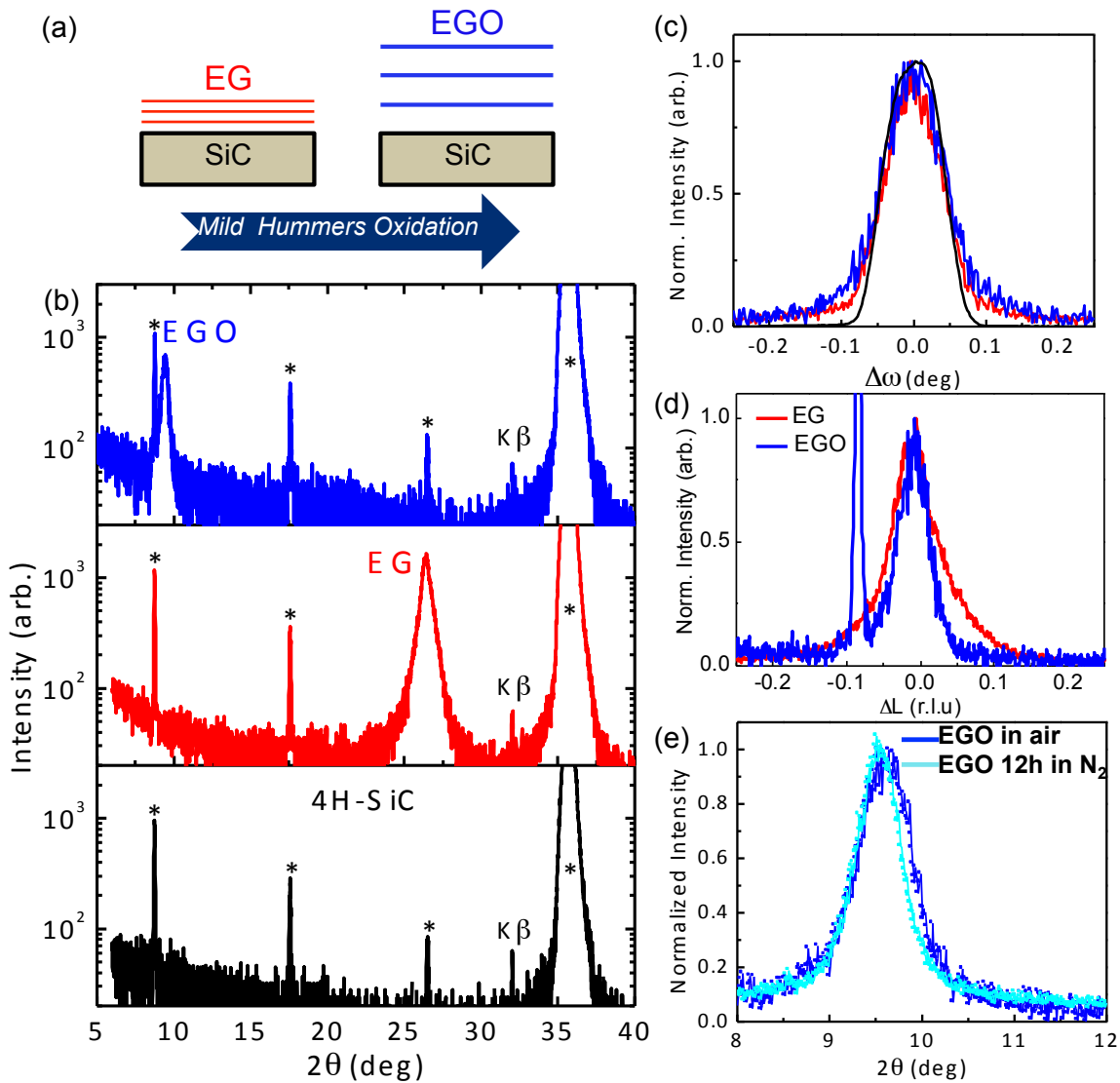

**Figure 1.** (a) Schematic illustration of the EG and EGO films. (b) XRD spectra of an 11(±1)-layer EGO film (blue solid line), a 12(±1)-EG film (red), and a bare 4H-SiC (000-1) substrate (black). The asterisks indicate peaks arising from the substrate. Peaks of EGO and EG are located at 2θ=9.46 deg and 2θ=26.42 deg, respectively. (c) Rocking curves (ω-scans) of EGO, EG, and the SiC substrate. These curves are, for comparison, centered around their mean value and normalized to the maximum peak intensity. The FWHM of the EGO, EG, and SiC peaks are 0.086 deg, 0.079 deg, and 0.073 deg, respectively. (d) Principal XRD peaks of EGO and EG centered around and normalized to its maximum value. For convenience, in these spectra the x-axis is expressed in reciprocal lattice units (r.l.u.) referred to the c-axis of 4H-SiC (c=10.08 Å). (e) Zoom in of the EGO peak at about 2θ=9.5 deg after keeping the sample in ambient conditions for approximately 2 weeks (dark blue), and after putting the same sample in a $N_2$ environment for one night (light blue).

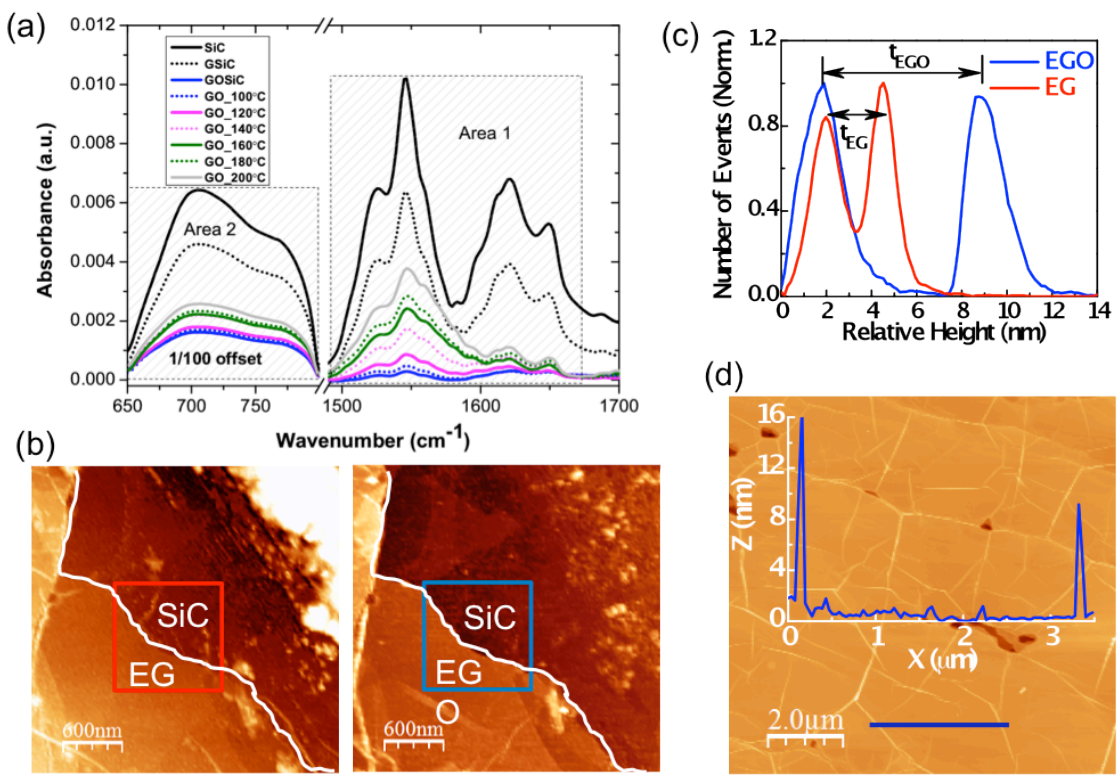

**Figure 2.** (a) Infrared absorbance spectra of bare SiC, EG (GSiC) and EGO (GOSiC/GO) at RT and after annealing at different temperatures. All these peaks are associated with silicon carbide and are consistently observed in the two samples: SiC and EG-on-SiC samples. (c) AFM topography images of an EG and EGO film on SiC. (c) Heights distribution in the squared regions shown in the topography images reported in panel (b); five-point smoothing was applied to obtain the red (EG) and blue (EGO) curves. See SI for more details. (d) AFM topography and height profile along the blue segment of the 12-layer EGO film used in the XRD measurements.

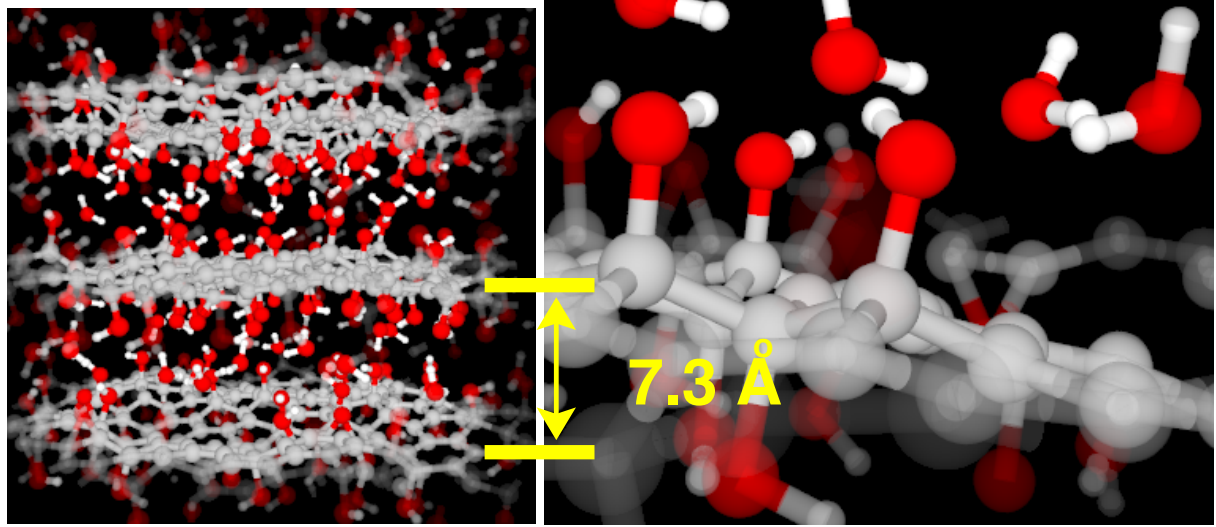

**Figure 3.** Model structure of EGO generated from DFT-D2 presenting the chemical composition reported in the table and a homogeneous lateral distribution of oxygen functional groups. Bottom-left panel, ball and stick illustration of the model structure of EGO; the periodic supercell includes 234 atoms and the relaxed structure shows an average interlayer spacing of 7.3 Å. Right panel, selected region of the model structure of EGO showing a large concentration of both hydroxyl groups and water molecules; the spatial distribution of hydroxyl groups is not compact and organized enough to form mechanically stable traps for water molecules bridging and holding nearest neighbor layers at distances larger than 9 Å.

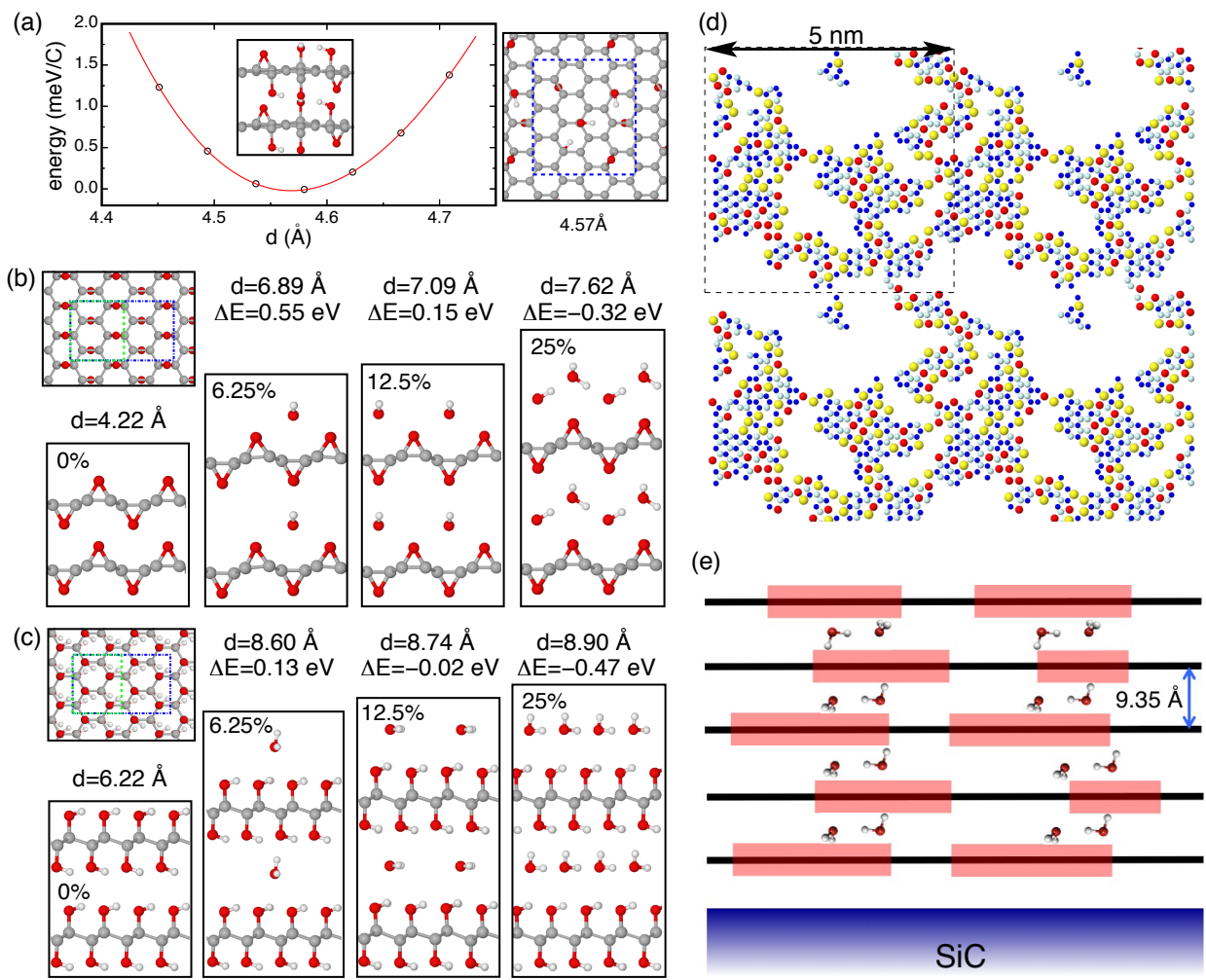

**Figure 4.** (a) Energy vs. interlayer spacing (symbols) values obtained from DFT-D2 calculations of a EGO model presenting a selected distribution of epoxide and hydroxyl species on the carbon layer, and no water molecules. Ball and stick illustrations of the model structure of EGO (inset and top view of a graphene oxide layer on the right side) are also shown. (b) Model structures of EGO used to mimic regions of the multilayer film where the graphene oxide layers are fully oxidized with epoxide groups and host an increasing concentration of water molecules in between. Colored rectangles in the top-left corner show the planar dimensions of the supercells used to models water concentrations of 6.25% (blue dash line) and 0%, 12.5%, and 25% (green dash line). For each model structure of EGO, the optimal interlayer spacing computed from DFT-D2 is reported on top of the ball-and-stick illustrations. In the case of the hydrated EGO models, the (zero-temperature and zero-pressure) enthalpy difference per water molecule

(ΔE) between hydrated and dry structures is also reported. ΔE is obtained by referring the energy of the hydrated model to that ones of a water molecule and the dry model. (c) Same as (b) with hydroxyl instead of epoxide species. (d) Graphene oxide layer presenting an inhomogeneous oxidation at the nanoscale of the carbon network (not shown) with epoxide (red and yellow colored balls) and hydroxyl (blue and cyan colored balls) chemisorbed on both sides of graphene (red and blue, one side, yellow and cyan, the opposite side). The model mimics the structure of a EGO film aged at room temperature; O:C ratio of 0.38, and fractions of hydroxyl and epoxide species equal to 0.27 and 0.11, respectively[17]. To mimic ageing and generate the model structure, we used the methods reported in Ref.[25]. (e) Schematic representation of a multilayer EGO film consisting of non-homogenously oxidized (red regions) and wetted graphene layers. The oxidized areas are rich in hydroxyl groups, while water molecules are trapped and form the contact between oxidized regions of neighboring layers.